\documentclass[prd,twocolumn,preprintnumbers,amsmath,amssymb,superscriptaddress,showpacs,showkeys,nofootinbib]{revtex4-1}
\usepackage{color}
\usepackage{amsmath,amsfonts,amssymb,amsthm,amstext,amscd,eucal,srcltx,mathrsfs}
\usepackage{epsfig,graphicx,bm}
\usepackage{epstopdf, epsf}
\usepackage{dcolumn}
\usepackage{hyperref}

\newcommand{\beq}{\begin{equation}}
\newcommand{\eeq}{\end{equation}}
\newcommand{\bea}{\begin{eqnarray}}
\newcommand{\eea}{\end{eqnarray}}
\newcommand{\beal}{\begin{aligned}}
\newcommand{\eeal}{\end{aligned}}

\newcommand{\mpl}{m_{\rm p}}
\newcommand{\lp}{\ell_{\rm p}}
\newcommand{\gn}{G_{\rm N}}

\begin{document}
\title{Lower dimensional corpuscular gravity and the end of black hole evaporation}
\author{Roberto Casadio}
\email{casadio@bo.infn.it}
\affiliation{Dipartimento di Fisica e Astronomia, Alma Mater Universit\`a di Bologna,
via~Irnerio~46, 40126~Bologna, Italy}
\affiliation{I.N.F.N., Sezione di Bologna, IS FLAG, viale~B.~Pichat~6/2, I-40127 Bologna, Italy}
\author{Andrea Giusti}
\email{agiusti@bo.infn.it}
\affiliation{Dipartimento di Fisica e Astronomia,
Alma Mater Universit\`a di Bologna,
via~Irnerio~46, 40126~Bologna, Italy}
\affiliation{I.N.F.N., Sezione di Bologna, IS FLAG, viale~B.~Pichat~6/2, I-40127 Bologna, Italy}
\affiliation{Arnold Sommerfeld Center, Ludwig-Maximilians-Universit\"at,
Theresienstra{\ss}e 37, 80333 M\"unchen, Germany}
\author{Jonas Mureika}
\email{jmureika@lmu.edu}
\affiliation{Department of Physics, Loyola Marymount University, Los Angeles, California, USA}
\begin{abstract}
Black holes in  $d < 3$ spatial dimensions are studied from the perspective of the corpuscular
model of gravitation, in which black holes are described as Bose-Einstein condensates of
(virtual soft) gravitons.
In particular, since the energy of these gravitons should increase as the black hole
evaporates, eventually approaching the Planck scale, the lower dimensional cases could
provide important insight into the late stages and end of Hawking evaporation.
We show that the occupation number of gravitons in the condensate scales holographically
in all dimensions as $N_d \sim \left(L_d/\lp\right)^{d-1}$, where $L_d$ is the relevant length
for the system in the $(1+d)$-dimensional space-time.
In particular, this analysis shows that black holes cannot contain more than a few gravitons
in $d=1$.  Since dimensional reduction is a common feature of many models of quantum
gravity, this result can shed light on the end of the Hawking evaporation.
We also consider $(1+1)$-dimensional cosmology in the context of corpuscular gravity,
and show that the Friedmann equation reproduces the expected holographic scaling as in higher
dimensions.
\end{abstract}
\maketitle
\section{Introduction}
The corpuscular description of gravity was first put forward in order to address some of the
fundamental issues which arise when considering quantum effects in black hole space-times~\cite{DvaliGomez},
and was subsequently applied to gravitation in the whole Universe~\cite{ads,cmb,roberto1,inflation,andrea}.
Dvali and Gomez~\cite{DvaliGomez} have proposed a new perspective in which black holes are viewed
as Bose-Einstein condensates (BEC) made of a very large number of soft 
(self-bound) gravitons.
The key observation in this picture is that the typical energy scale $\varepsilon_{\rm G}$ of the
gravitons is very small in most physically relevant systems and the weak field approximation should yield
reliable results.
\par
One can estimate $\varepsilon_{\rm G}$ by noting that the gravitational potential of a localised source of mass $M$
and radius $R$ is reproduced by a coherent state of $N_{\rm G}$ gravitons of wavelength
$\lambda_{\rm G}\sim R$~\cite{muck,m&gL,m&gA}, where~\footnote{We shall use units with $c=1$,
the four-dimensional Newton constant $\gn=\lp/\mpl$, where $\lp$ and $\mpl$ are the Planck length and mass,
respectively, and $\hbar=\lp\,\mpl$.}
\beq
N_{\rm G}
\simeq
{M^2}/{\mpl^2}
\ ,
\label{kaupL}
\eeq
and the typical single graviton energy~\footnote{Eq.~\eqref{EgN} follows from the Compton relation and can be
given a more precise meaning by including the (``post-Newtonian'') graviton self-interaction~\cite{m&gL,m&gA}.} 
\beq
\varepsilon_{\rm G}
\simeq
{\hbar}/{\lambda_{\rm G}}
\simeq
\mpl\,{\lp}/{R}
\ ,
\label{EgN}
\eeq
which, of course, sets an extremely small scale for astrophysical sources.
The total energy stored in the gravitational field is then well approximated by the Newtonian expression,
\beq
N_{\rm G}\,\varepsilon_{\rm G}
\simeq
N_{\rm G}\,\frac{\hbar}{\lambda_{\rm G}}
\simeq
M\,\frac{R_{\rm H}}{R}
\simeq
\frac{\gn\,M^2}{R}
\ ,
\label{Ug}
\eeq
where $N_{\rm G}\,\varepsilon_{\rm G}\ll M$ for any regular astrophysical object of size
$R\gg R_{\rm H}=2\,\gn\,M$.
However, if the object collapses to form a self-gravitating black hole, Eq.~\eqref{kaupL} yields the Kaup
limit~\cite{kaup} if the energy contribution of baryonic matter becomes subleading and the total ADM
mass of the system is mostly provided by the gravitons themselves~\cite{DvaliGomez,cgmo}, that  is
\beq
N_{\rm G}\,\varepsilon_{\rm G}
\sim
M
\ .
\eeq
At this stage the gravitons superpose to form a critical BEC and their Compton length, from Eq.~\eqref{Ug},
becomes
\beq
\lambda_{\rm G}
\simeq
R_{\rm H}
\simeq
\sqrt{N_{\rm G}}\,\lp
\ .
\label{Lrh}
\eeq
Nonetheless, it remains true that the single graviton energy
\beq
\varepsilon_{\rm G}
\simeq
{\mpl}/{\sqrt{N_{\rm G}}}
\ ,
\label{Eg}
\eeq
is still extremely small for an astrophysical black hole, and the weak field approximation should 
yield a reliable quantum description of such objects. 
\par
A clear merit of the corpuscular picture is that the horizon area scales holographically,
since Eq.~\eqref{Lrh} implies that
\beq
A_{\rm H}
=
4\,\pi\,R_{\rm H}^2
\simeq
N_{\rm G}\,\lp^2
\ .
\label{Ah}
\eeq
A less obvious consequence is that the Hawking evaporation is not a vacuum process,
but it is rather due to the $2 \to 2$ scattering of gravitons inside the condensate which yields
a total depletion rate 
\beq
\hbar\,\Gamma
\sim
N_{\rm G}^2\,\alpha_{\rm G}^2\,\varepsilon_{\rm G}
\simeq
{\mpl}/{\sqrt{N_{\rm G}}}
\ ,
\label{eq:Grate}
\eeq
where the interaction strength $\alpha_{\rm G}\sim \varepsilon_{\rm G}^2/\mpl^2\sim N_{\rm G}^{-1}$.
The number of gravitons in the BEC will then decrease according to~\cite{DvaliGomez}
\beq
\dot N_{\rm G}
\simeq
-\Gamma
\simeq
-\frac{1}{\sqrt{N_{\rm G}}\, \lp} 
\ .
\label{dotN}
\eeq
From Eq.~\eqref{kaupL}, one then obtains
\beq
\dot M
\simeq
\frac{\mpl\,\dot N_{\rm G}}{\sqrt{N_{\rm G}}}
\simeq
-\frac{\mpl^3}{\lp\,M^2}
\ ,
\label{dotM}
\eeq
which reproduces the famous Hawking behaviour. 
It is however important to remark that the BEC of gravitons has no thermal character and
the corresponding Hawking temperature arises as an effective classical feature~\cite{DvaliGomez}.  
\par
It is also important to note that, as the black hole evaporates and $N_{\rm G}$ decreases,
the typical graviton energy~\eqref{Eg} instead increases, eventually approaching the Planck
scale near the end of the evaporation.  
It has been conjectured that scattering processes at such high energies should effectively probe
a lower number of spatial
dimensions~\cite{tHo93,Mann:1991qp,rbm11,mann2,mann3,btz,Casadio:2015jha,rmjmentropic,jmpnepj,jrmpbh}.
Hence, it appears natural to investigate a corpuscular description for a
(relatively) small number $N_{\rm G}$ of gravitons in spaces of dimension $d<3$ in order to gain
some insight about late stages and the end of the evaporation.  Indeed, we expect higher-order corrections
to Eq.~\ref{dotM} of the form $1/N^{3/2}$ \cite{DvaliGomez}, and therefore we view the dimensional reduction
as an effective way of describing these.
Of course, we must warn the reader that any attempt at extrapolating low-energy frameworks (like the
Hawking effect and the corpuscular picture of very large black holes) all the way into the Planck scale
is inherently hazardous without assuming a specific theory of quantum gravity.
\par
Before we proceed, we stress that the gravitons in the black hole condensate are bound in their own (mean field)
potential~\cite{QHBH} and, of course, they do not propagate unless they scatter off and deplete (a more detailed
description of the ground state and Hawking radiation can be found, e.g.~in Refs.~\cite{Tbh}).
This picture is similar to the Coulomb potential viewed as made of virtual photons~\cite{muck1}, whose
number is indeed not directly observable, but can be used to compute the electrostatic potential energy.
At the same time, we remark that we do not consider the lower dimensional cases as a fundamental
picture of the space-time, but just as an effective way to describe the degrees of freedom of such virtual gravitons
as the evaporation proceeds to very small black hole masses and the energy of the virtual gravitons
approaches the Planck scale.
\section{A review of $(1+1)$-dimensional gravity}
The characteristics of gravitation in space-times of dimension less than four are quite unusual,
and we shall recall just a few results of interest for the present work, including the dimensionality
of the Newton constant $G_d$ in $(1+d)$ dimensions.
\par
formulations of a gravitational theory in $(1+1)$ dimensions have been considered for decades,
initially as a pedagogical curiosity, and later as probes of quantum gravity
effects~\cite{gott1,gott2,jackiw,rbm11,robb2,grumiller}. 
The corresponding models require the inclusion of an additional scalar field  (the dilaton)
in the action, which in most cases introduces additional physics in the gravitational sector.
This is in contrast to Einstein gravity in $(3+1)$ dimensions, which obviously requires no such auxiliary
field.
\par
There is one formulation of two-dimensional gravity that stands apart, however.
The ``$R=T$'' Liouville formulation (whose metric solution was discussed in Refs.~\cite{rbm11},
and later expanded upon in {\it e.g.}~\cite{mann2,mann3}) has the unique attribute that the dilaton
completely decouples from the gravitational sector which, along with the resulting form of the metric
in~(\ref{g11}), makes it the closest analogue to pure Einstein gravity in lower dimensions.
The action is
\beq
S_2
=
\int 
d^2x \,\sqrt{-g}
\left\{\frac{1}{16\,\pi\,G_1}
\left[\psi\, R
+\Lambda
+\frac{1}{2}(\nabla \psi)^2
\right]
+{\cal L}_m
\right\}
\ ,
\label{s2dilaton}
\eeq
where $\psi$ is the aforementioned dilaton.
Varying (\ref{s2dilaton}) with respect to $\psi$ and $g_{\mu\nu}$ gives
the respective field equations
\beq
 R
 =
 \Box\psi
 \label{vareq1}
 \eeq
and 
\bea
\frac{1}{2}\,\nabla_\mu \psi \,\nabla_\nu \psi
- g_{\mu \nu} \left(\frac{1}{4}\,\nabla^\lambda \psi \,\nabla _\lambda \psi - \Box \psi\right)
-\nabla_\mu \nabla_\nu \psi
&&
\nonumber
\\
=
8\,\pi\, G_1\, T_{\mu \nu}
+ \frac{\Lambda}{2}\, g_{\mu \nu}
\ ,
\qquad
&&
\label{vareq2}
\eea
where the stress-energy tensor is defined by
\beq 
T^{\mu\nu}
=
-\frac{2}{\sqrt{-g}}\,\frac{\delta {\cal L}_m}{\delta g_{\mu\nu}}
\ .
\eeq
By taking the trace of the second field equation (\ref{vareq2}), one finds the left-hand side recovers (\ref{vareq1}),
which removes the dilation from the field equations.
This reduces (\ref{vareq2}) to the standard Liouville gravity form
\begin{equation}
R-\Lambda
=
8\,\pi\, G_1\, T
\ .
\label{eq2}
\end{equation}
Note that the exact form of $\psi$ is irrelevant to the solution.
\par
For $\Lambda = 0$,  the solution to (\ref{eq2}) is~\cite{rbm11}
\beq
ds^2
=
\left(1-2\,G_1\, M\, r\right)
dt^2
-
\frac{dr^2}{\left(1-2\,G_1\, M\, r\right)}
\ ,
\label{g11}
\eeq
where $r> 0$ is the spatial coordinate.
Note this metric is a natural lower-dimensional form of the general $(4+n)$-dimensional Schwarzschild solution
\beq
f(r) = 1-\frac{2G_n M}{r^{n+1}}
\eeq 
in the case $n=-2$ ({\it i.e.} 2-D spacetime), which adds to its appeal over other dilaton models. 
\par
The (point) mass $M$ is defined in terms of the energy density $T^{00}$ as~\cite{jmpn11}
\beq
T^0_{\ 0}
=
\rho 
=
\frac{M}{2\,\pi\, G_1}\,\delta(r)
\ .
\eeq
It should be noted that the coupling $G_1$ is dimensionally different from the four-dimensional Newton
constant $G_3=\gn = \lp/\mpl$, which has central implications for quantum gravity and will be discussed shortly.  
\par
The metric~(\ref{g11}) gives the horizon radius as
\beq
R_1
=
\frac{1}{2\,G_1\, M}
\ ,
\label{2dh}
\eeq
which is the gravitational radius that sets the Compton length of interest in the corpuscular framework. 
Moreover, the Hawking temperature can readily be calculated from the surface gravity approach, and is
\beq
T
=
\frac{G_1\, M}{2\,\pi}
\ ,
\eeq
from which it follows that the black hole entropy is
\beq
S
=
\frac{2\,\pi}{G_1}\, \ln\left(\frac{M}{m_*}\right)
\,
\eeq
which is not holographic ({\it i.e.}~scaling as the horizon area) as with higher dimensions.
The logarithmic form necessitates the introduction of a new mass scale $m_*$, which 
may or may not be related to the Planck scale. These two-dimensional thermodynamic quantities have recently been shown to arise as the result
 of an effective dimensional reduction behaviour for black holes of sub-Planckian mass~\cite{bcjmpn1}.
\par
Alternate approaches to $(1+1)$-dimensional gravity retain the dilaton as a component of the field equations,
which consequently plays a role in the thermodynamics.
For example, in Ref.~\cite{grumiller}, the entropy is found to be entirely dependent on the value of the
dilaton at the horizon, {\it i.e.}
\beq
S
=
2\,\pi\, \psi_H
\ ,
\eeq
which can be re-expressed as
\beq
S
\sim
\frac{A_H}{G_{\rm eff}}
\ ,
\eeq
where $G_{\rm eff} = {G_1}/{\psi_H}$.
Noting that the area of a $(1+1)$-dimensional black hole consists of two antipodal points,
this gives the entropy as a constant of order $G_{\rm eff}^{-1}$. 
This is consistent with the result derived in Ref.~\cite{rmjmentropic}, where the entropy of 
a two-dimensional black hole is found to be $S = N \sim {\mathcal O}(1)$,
where $N$ is the number of holographic bits.
\par
The mass dependence of the horizon size in $d=1$ is special from the context of a quantum theory. 
According to Eq.~\eqref{2dh}, the horizon varies inversely with the mass. 
The coincidence between this and the usual quantum length scales is not by accident.
In fact, when one considers the Planck unit dependence of the associated Newton's constant $G_1$,
the association is clear.
The general expression for the $(1+d)$-dimensional Newton constant is~\cite{rmjmentropic}
\beq
G_d
=
2\,\pi^{1-\frac{d}{2}}\, \Gamma\left(\frac{d}{2}\right)
\frac{\ell_d^{d-1}}{\hbar}
\simeq
\frac{\ell_d^{d-1}}{\mpl\,\lp}
\ .
\label{newtond}
\eeq
where $\ell_d$ is the ($1+d$)-dimensional Planck length.
When $d=1$, (\ref{newtond}) is independent of this quantity
and it follows that
\beq
G_1  \sim \frac{1}{\mpl\,\lp}
\ .
\eeq 
That is, Newton's constant in $(1+1)$ dimensions is naturally ``dual'' to $\hbar$. 
\par
In fact, this suggests that the Compton and horizon scales are essentially indistinguishable quantities. 
This has previously been shown to be an artefact of the self-completeness prescription in $(1+1)$
dimensions~\cite{jmpnepj}.
It is furthermore supported by the fact that gravitation in $d=1$ does not admit a generalised uncertainty
principle~\cite{Casadio:2015jha}, 
\beq
(\Delta x)_{\rm Total} = (\Delta x)_Q + (\Delta x)_G
\label{gup}
\eeq
where $(\Delta x)_Q$ is the usual Heisenberg uncertainty and $(\Delta x)_G \sim R_{\rm H}$
is the gravitational uncertainty given by the horizon size.
In $d=1$, the horizon is given by Eq.~\eqref{2dh},
which is identically the Compton wavelength, and Eq.~\eqref{gup} reduces to
\beq
(\Delta x)_{\rm Total}
\sim
{\hbar}/{M}
\ .
\eeq
\section{Corpuscular Black Holes}
One of the primary results of the corpuscular model is that the gravitational entropy scales as the horizon area
in $d=3$,
\beq
S
\sim
N_{\rm G}
\sim
A_{\rm H}
\ ,
\eeq
as follows from Eq.~\eqref{Ah} for black holes, and a similar relation for the de~Sitter
space-time~\cite{ads,cmb,inflation}.
In particular, we remark here that the number of gravitons $N_{\rm G}$ can be obtained from the
total energy~\eqref{Ug} stored in the gravitational field,
\beq
U_3
\simeq
\frac{G_3\,M^2}{R}
\ ,
\label{U3}
\eeq 
provided the relation~\eqref{EgN} holds for $R=R_3\equiv R_{\rm H}$.
\par
In principle, the laws of quantum mechanics are independent of the space-time dimension in which they live.
Quantities such as the de~Broglie or Compton wavelengths define length scales that are characteristic of the object,
rather than the environment.
The primary ingredient to be considered in dimensional reduction scenarios is therefore the gravitational contributions.
In the following, we will thus assume the above scheme continues to hold as the black hole evaporates,
$\varepsilon_{\rm G}$ becomes large and the dynamics is described by lower dimensional gravity.
In practice, we shall always estimate the single graviton energy $\varepsilon_{\rm G}$ according to
Eq.~\eqref{EgN}, with $R=R_d$ for a black hole, $R_d$ being the horizon radius in $d$ spatial dimensions,
and obtain $N_{\rm G}=N_d$ from the lower dimensional versions of Eq.~\eqref{U3}.
\par
In $d=1$, the total gravitational binding energy is
\beq
U_1
\simeq
G_1\,M^2\,R
\ .
\label{2du}
\eeq
This can be derived from the weak-field limit of the metric, and is also consistent with the equations of motion
of a test particle in the space-time~\cite{rbm11}.
It is also consistent with the gravitational force between masses being constant in $d=1$~\cite{rmjmentropic}.
\par
Since the entropy does not follow a holographic law, it is not completely intuitive how to relate the occupation
number $N_{\rm G}$ to these quantities from thermodynamical considerations.  
We notice, however, that we can write
\beq
U_1
\simeq
\frac{M R}{R_1}
\ ,
\eeq
where we used Eq.~\eqref{2dh}, from which
\beq
N_1
\simeq
\frac{U_1}{\varepsilon_{\rm G}}
\simeq
\frac{R^2 M^2}{\lp^2\, \mpl^2}
\ .
\label{1dn}
\eeq
Note now that for a black hole with $R = R_1$, this simply becomes $N_1 \simeq 1$, 
which is the expected holographic behavior in $d=1$.
\par
Comparing this to the $d=3$ case, the occupation number is of order unity only when the mass (and size)
of the object are Planckian, {\it i.e.}
\beq
N_1
\sim
\frac{M^2}{\mpl^2}
\simeq
\frac{R^2}{\lp^2}
\simeq
1
\ ,
\eeq
suggesting that $d=1$ black holes are strictly Planckian objects in the BEC picture.
Moreover, this indicates that $d=1$ black holes consist of (roughly) a single graviton, supporting the
notion that such two-dimensional black holes are fundamentally quantum mechanical objects.
Finally, we note that the premise of classicalization, {\it i.e.} requiring $N \gg 1$, is no longer valid,
which further supports the full quantum nature of a two-dimensional BH.
Incidentally, this result is also similar in spirit to the bit count (density) derived in Ref.~\cite{rmjmentropic},
which concluded there is one bit of information at every point in the $(1+1)$-dimensional space-time.
\section{Thermodynamics and late decay}
Since we have shown that a $d=1$ BEC black hole consists of a few gravitons,
it now makes no sense to describe it as a thermodynamic object (and most likely unstable,
see~\cite{planckBH} and references therein).
In the classical theory, however, such black holes have a well-defined temperature and
entropy as described above.
We have already shown in Eq.~\eqref{dotM} that an effective thermal description arises
in the corpuscular picture of four-dimensional black holes from the depletion of the BEC,
and we can here repeat the steps in order to recover a corresponding ``temperature'' 
in the $(1+1)$-dimensional case.
We remark once again that we view this lower dimensional model as the limiting case of 
the evaporation in $d=3$, assuming that, in the dimensional transition, the black hole begins
to take some (but not all) two-dimensional characteristics.
\par
Specifically, the emission rate can be calculated to obtain an expression for the BEC
black hole temperature.
In terms of the occupation number $N_{\rm G}\simeq N_1$, this can be estimated as the scattering
cross-section of two soft gravitons, $\Gamma \sim {N_1^2\, \alpha_1^2}/{R}$.
The characteristic wavelength of the system can be expressed in terms of $N_{\rm G}$
and $M$ by using Eq.~(\ref{1dn}), that is $R \sim \lp\,\sqrt{N_1}\,{\mpl}/{M}$,
and thus the decay rate is
\beq
\Gamma
\sim
\frac{N_1^2\, \alpha_1^2\, M}{\sqrt{N_1}\,\lp\,\mpl}
\ .
\eeq
Normally, the coupling constant $\alpha_{\rm G}$ is related to the occupation number
by the maximal packing relation $\alpha_{\rm G}\, N_{\rm G} \simeq 1$~\cite{DvaliGomez}.
To address the value of $\alpha_1$, we approach the problem from the perspective
that in order to transition to a $d=1$ state, the black hole must first start in the $d=3$
regime (we ignore $d=2$ due to the requirement of AdS space).
In particular, we assume the black hole is ``almost'' in its final $d=1$ state, but still retains
the thermal character of the $(3+1)$-dimensional space-time during the transition.
We therefore set the coupling as $\alpha_1 \sim {1}/{N_{\rm G}}$,
and the decay rate becomes
\beq
\Gamma
\sim
\frac{M}{\sqrt{N_{\rm G}}\,\lp\,\mpl}
\ .
\eeq
The radiative thermal power in $d$ spatial dimensions is given by
$\dot M \sim -A_{d-1}\,T^{d+1}$, where $A_{d-1} \simeq R_{\rm H} ^{d-1}$
is the horizon area~\cite{jrmpbh}.
In the usual case of a $d=3$ black hole, the area $A_3\simeq R^2 \sim M^2$ and
the temperature $T \sim M^{-1}$, so the standard Hawking mass loss is
\beq
\dot M
\sim
-M^{-2}
\sim
-T^2
\ .
\label{standardH}
\eeq
For a $d=1$ black hole, this is modified.
First, the area of such a black hole consists of the two antipodal points that constitute
the horizon, and so $A_1$ is substantially constant.
Since the temperature scales linearly with the mass in $d=1$, the rate is thus
\beq
\dot M
\sim
-T^2
\sim
-M^2
\ .
\label{standardH1}
\eeq
\par
The derivative $\dot M=d M/d t$ can be estimated in terms of the decay rate by noting the loss in mass
$dM \sim -\hbar/\lambda_{\rm e}$, where $\lambda_{\rm e}$ is the threshold wavelength for escaping
from the condensate, occurs in a time $dt \sim \hbar\, \Gamma^{-1}$.
Combining these and setting $\lambda_{\rm e} \simeq \lambda_{\rm G}\simeq  {\lp\, \mpl}/{M}$, we obtain
\beq
\frac{d M}{dt}
\simeq
-
\frac{M^2}{\sqrt{N_{\rm G}}\,\lp\,\mpl}
\ .
\eeq
Since $\sqrt{N_{\rm G}}\,\mpl\simeq M$, this does not quite look like the expected temperature profile.
However, noting that in the final limit $N_{\rm G}=N_1\simeq 1$ for a black hole in $d=1$, 
one can ultimately conclude that the corpuscular approach agrees with the (semi-)classical
result~\eqref{standardH1}.
%
%
%
%
%
\section{Corpuscular $(1+1)$-D Cosmology}
We note for completeness that the lower dimensional version of the BEC formalism may also be applied
consistently to cosmology.  
In $d=3$, a cosmological BEC was shown to also obey a holographic relationship~\cite{ads,roberto1}.
In particular, integrating over the entire Hubble volume $L_3^3\sim H_3^{-3}$ the Friedman equation
$H_3^2 \sim G_3\, \rho$, one finds $L_3 \simeq G_3\, M_3$, which is the same relation for a black hole
horizon.
It is straightforward to check and see if the same is true for $d=1$ gravity.
From the Friedman equation
\beq
H_1^2
\sim
G_1\, \rho
\ ,
\eeq
we have $H_1 \sim L_1^{-2}$, where $L_1$ is the one-dimensional Hubble length,
and $\rho \sim M_1/L_1$ is the linear energy density of the cosmological BEC.
Integrating over the length of the space-time, we obtain
\beq
L_1^{-1}
\sim
G_1\, M_1
\ ,
\eeq
which looks exactly like the horizon relation~\eqref{2dh}.
If we assume there are $N_1$ gravitons of wavelength $L_1$ comprising the mass,
then we can replace $M_1 \sim {\hbar}/{L_1}$, and we find
\beq
L_1^{-1}
\sim
G_1\, N_1\, {\hbar}\,L_1^{-1}
\ .
\eeq
Recalling that $G_1 = \hbar^{-1}$ and solving for $N_1$, we conclude again that
$N_1 \sim 1$, {\it i.e.}~the relationship is holographic for the $d=1$ space.
However, care should be take to avoid concluding that this says there is exactly only one graviton
in the cosmological BEC, since other constant factors have been omitted for the sake of brevity.
\par
In $d=2$, it is necessary to consider the cosmological BEC formalism similar to that in~\cite{roberto1},
due to the fact that it is well-known that gravity in this number of space-time dimensions is topological,
and has no propagating degrees of freedom~\cite{carlip1}.
This implies there are neither gravitational waves, nor quantum excitations ({\it i.e.}~gravitons).
This, however,  does not necessarily preclude a corpuscular description.
A solution to Einstein's equations in $d=2$ for a black hole of mass $M$ and angular momentum $J$
is the famed BTZ metric~\cite{btz}
\bea
ds^2
&=&
\left(G_2\, M - \frac{r^2}{L^2}-\frac{J^2}{4\,r^2}\right) dt^2
-\frac{dr^2}{G_2\, M - \frac{r^2}{L^2}-\frac{J^2}{4\,r^2}}
\nonumber
\\
&&
-r^2\left(d\phi - \frac{J}{2\,r^2}\;dt\right)^2
\ ,
\label{btzg}
\eea
where $L^{-2}\sim -\Lambda_2$ is the scale that defines the cosmological constant.
In absence of spin, this gives rise to black holes with horizon for AdS spaces only,
\beq
R_2
=
{\sqrt{G_2\,M}}\,{L}
\ ,
\label{threedh}
\eeq
and we note that the dimension of the gravitational constant is $G_2 \simeq \mpl^{-1}$.
\par
We can, however, focus on the cosmological BEC to derive a scaling dependence on $N_2$,
and based on the findings of the previous discussions, assume that the same will hold for the black hole case.
In fact, the BTZ black hole solution is potentially interesting from the point of view of the BEC formalism
because the horizon depends on both the quantum and cosmological length scales.
As discussed in~\cite{roberto1}, the cosmological constant may itself be a manifestation of a universal
BEC that obeys holographic behavior in the same way as quantum gravity.
\par
In the absence of pressure and spatial curvature, the relevant Friedman equation in $d=2$
is given by~\cite{ndfriedman} 
\beq 
H_2^2 \sim G_2\,\rho
\ .
\eeq
Since $H_2 \sim L_2^{-1}$, integrating the above over the area enclosed by $L_2$ gives
\beq
1 \sim G_2\, \rho\,L_2^2
\quad
\Longrightarrow
\quad
1 \sim G_2\, M_2
\ .
\eeq
As in~\cite{roberto1}, the above expression gives a horizon-like relation for a cosmological mass scale
$M_2 =\rho\,L_2^2$, except in this case it is a saturation relation.
Replacing $M_2 = N_2\, \epsilon_2$, where $\epsilon_2 \sim {\lp \mpl}/{L_2}$, this yields
\beq
\lp\,N_2 \sim L_2
\label{d2n}
\eeq
and so the occupation number scales linearly with size.
Although this differs from the scaling relation in $d=3$, it still conforms to a holographic relation
if one considers that in $d=2$, the horizon area is one-dimensional, {\it i.e.}~a perimeter.
Another interesting observation is that this occupation number does not depend on the black hole mass,
but rather is a constant that is dependent on the largest scale size of the space-time.

\section{Conclusions}
To conclude, we have considered the end stage of black hole evaporation by extending the BEC
formalism to $(1+1)$-dimensional space-time.
We find that the corpuscular description is consistently holographic for the appropriate space-time
dimension, and furthermore interpret our results to imply that such black holes consist of a single
graviton.
We can nevertheless recover the standard thermal portrait of such $(1+1)$-dimensional black holes by
considering the evaporation as the result of graviton-graviton scattering in the BEC in the near-end
evaporation limit.
We also show that the holographic nature of lower-dimensional gravitation can be similarly derived
by assuming the existence of a cosmological BEC responsible for {\it e.g.}~dark forces.
This is particularly relevant to the $(1+2)$-dimensional case, in which a Newtonian limit does not
exist and a corpuscular black hole description is not well-defined.
\par
Finally, it is important to stress that the process of dimensional reduction for self-gravitating systems
is still one of the major open problems for the theory of gravitation in lower dimensions.
Indeed, it is still unclear whether this dimensional reduction would dynamically emerge as a discrete
jump (or jumps at various scales) or as a continuous transition involving intermediate fractional dimensions
(see e.g.~\cite{Calcagni:2011kn,Diaz:2017hdd}).
\begin{acknowledgments}
R.C.~and A.G.~are partially supported by INFN, research initiatives FLAG, and
their work has been carried out in the framework of GNFM and INdAM and the COST
action {\em Cantata\/}.
J.M.~thanks the Department of Physics and Astronomy of Bologna University and
INFN for their generous hospitality during the initial stages of this project.
\end{acknowledgments}
\end{document}